\documentclass[aps,pre,twocolumn,groupedaddress]{revtex4-1}

\usepackage{graphicx}
\usepackage{amssymb}
\usepackage{amsmath}
\usepackage[usenames,dvipsnames]{xcolor}

\DeclareMathOperator\erfc{erfc}
\begin{document}


\title{Humidity-insensitive water evaporation\\ from molecular complex fluids}

\author{Jean-Baptiste Salmon}
\email[]{jean-baptiste.salmon-exterieur@solvay.com}
\affiliation{CNRS, Solvay, LOF, UMR 5258, Univ. Bordeaux, F-33600 Pessac, France.}

\author{Fr\'ed\'eric Doumenc}
\affiliation{Laboratoire FAST, Univ. Paris-Sud, CNRS, Universit\'e Paris-Saclay, F-91405, Orsay, France,}
\affiliation{Sorbonne Universit\'es, UPMC Univ Paris 06, UFR 919, F-75005 Paris, France}

\author{B\'eatrice Guerrier}
\affiliation{Laboratoire FAST, Univ. Paris-Sud, CNRS, Universit\'e Paris-Saclay, F-91405, Orsay, France.}
\date{\today}

\begin{abstract}
We  investigated theoretically water evaporation from
concentrated supramolecular mixtures, such as solutions of polymers or
amphiphilic molecules, using numerical resolutions of a one dimensional model based on mass transport equations. Solvent evaporation leads to the formation of a concentrated solute layer at the drying interface, which slows down evaporation 
in a long-time scale regime. In this regime, often referred to as the {\it falling rate period}, evaporation is dominated by diffusive mass transport within the solution, as already known. 
However, we demonstrate that, in this regime, the rate of evaporation does not also depend on the ambient humidity 
for many molecular complex fluids. Using analytical solutions in some limiting cases, 
we first demonstrate that a sharp decrease of the water chemical activity at high solute 
concentration, leads to evaporation rates which depend weakly on the  humidity, as 
the solute concentration at the drying interface slightly depends on the humidity. However, 
we also show that a strong  decrease of the mutual diffusion coefficient of the solution enhances considerably this effect, leading to nearly independent evaporation 
rates over a wide range of humidity. The decrease of the mutual diffusion coefficient indeed
induces strong concentration gradients at the drying interface, which {\it shield}
the concentration profiles from humidity variations, except in a very thin region close to the drying interface.

\end{abstract}


\maketitle

\section{Introduction}
\begin{figure}[ht]
\begin{center}
\includegraphics{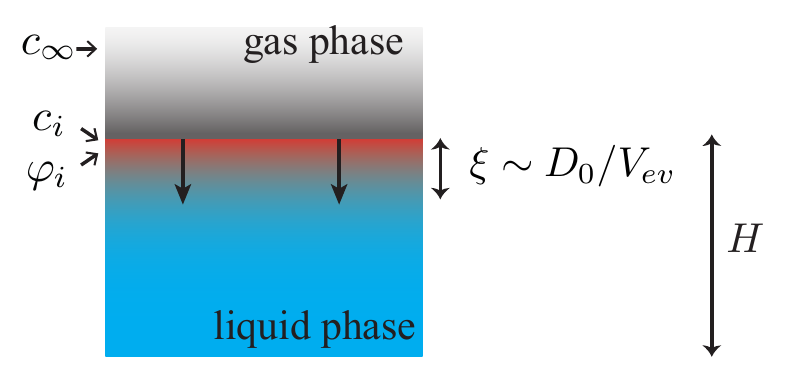}
\caption{Evaporation of a thick film. Water evaporation from the liquid binary mixture induces a receding of the drying interface at a rate $V_{ev}$ (m/s). Colors code for the solute concentration within the fluid, whereas the grayscale codes for the mass transfer within the gas phase. $\xi \sim D_0 / V_{ev}$ is the typical scale of the evaporation-induced concentration
polarization layer, $c_{\text{i}}$ the water vapor concentration at the drying interface, $c_{\infty}$ the water concentration in surrounding ambient air,
and $\varphi_i$ the solute concentration within the fluid at the drying interface. 
\label{fig:Sketch1}}
\end{center}
\end{figure}

Water evaporation from binary molecular complex fluid solutions (polymers, amphiphilic molecules, etc.) is a common feature of many different experimental situations ranging from
the drying of water-borne polymeric coatings to spray-drying in food engineering~\cite{Mujumdar}. For ambient conditions and slow drying, 
 water in the solution is at equilibrium with its vapor at the liquid-gas interface, 
and evaporation is driven by the 
vapor mass transfer towards the surrounding air, see Fig.~\ref{fig:Sketch1} 
for the case of the drying of a thick film~\cite{Bird,Cussler}.
For low volatile solvents such as water at ambient conditions, the 
rate of evaporation  from the drying interface takes then the following form: 
 \begin{eqnarray}
\rho V_{ev} = k (c_{\text{i}} - c_{\infty }),
 \end{eqnarray}
 where $\rho $ is the  density of liquid water, $k$ a mass transfer coefficient (m/s), $c_{\text{i}}$ the water vapor concentration at the interface, and $c_{\infty} $ the water vapor concentration in the surroundings~\cite{Bird,Cussler}. For the slow drying configurations considered here, we assume isothermal conditions. 
At early time scales and for dilute solutions, the low solute concentration within the bulk fluid hardly affects the water chemical activity, leading thus to:
 \begin{eqnarray}
\rho V_{ev} = k c_{\text{sat}} (1 - a_e),
 \end{eqnarray}
 where 
$c_{\text{sat}}$ (kg/m$^3$) is the concentration at saturation of water in the vapor phase for pure water, and
$a_e$ the  humidity of the ambient air~\cite{Bird,Cussler}.
In this regime, also known as the {\it constant rate period} in the context of 
polymer coatings~\cite{Saure98,Guerrier:98,Okazaki:74}, the evaporation rate is nearly constant, and 
depends on both the ambient humidity (term $a_e$) and mass transfer within the gas phase (term  $k$). 
Water evaporation, in turn, drives also  a receding of the drying interface, at a rate $V_{ev}$  owing to mass conservation.  The receding of the interface  consequently concentrates the non-volatile solutes at the air-solution interface in a layer
of thickness  $\xi \sim D_0 / V_{ev}$, where $D_0$ is the solute diffusivity in the dilute regime, see Fig.~\ref{fig:Sketch1}.
 Such a layer, also known as the {\it concentration polarization layer} in the field of membrane science~\cite{Belfort:94},  arises from the competition between diffusion and the displacement of the interface, a common feature of many different processes such as ultra-filtration or drying. 

In the following, we focus on experimental situations for which the polarization layer remains smaller than 
the total thickness $H$ of the solution, i.e. $H \gg D_0/V_{ev}$. Such a regime is not only key for understanding the drying of sessile drops~\cite{Kajiya:06,Pauchard:03,Baldwin:11} or of thick films~\cite{Okunozo:06,Okuzono:08}, but also for investigating 
evaporation-driven water transport through concentrated supramolecular solutions~\cite{Spaar:00,Roger:16} (see below).
At long time scales, accumulation of solutes in the polarization layer also affects the water chemical activity at the air/fluid interface, and the latter is now at equilibrium with air saturated with water at a concentration $c_i = a(\varphi_i) c_{\text{sat}}$, where $\varphi_i$ is the solute concentration at the drying interface, and $a(\varphi_i)$ the corresponding water chemical activity. Solute accumulation thus slows down the drying kinetics as the evaporation 
rate now follows:
\begin{eqnarray}
\rho V_{ev} = k c_{\text{sat}} (a(\varphi_i)-a_e). \label{eva2} 
 \end{eqnarray}
In this regime, also known as the {\it falling rate period} in the field of polymer coatings~\cite{Saure98,Guerrier:98,Okazaki:74}, 
$\varphi_i$ increases asymptotically towards $\varphi^\star$ given by the local chemical equilibrium $a(\varphi^\star) =  a_e$, and  
$V_{ev} \to 0$ leading to a broadening of the concentrated layer as the latter evolves as $\xi \sim D_0/V_{ev}$.
In this long time scales regime, the evaporation rate does not depend anymore on mass transfer within the vapor phase, but only on solute diffusion through the polarization layer~\cite{Saure98,Guerrier:98,Okazaki:74}. 
One still nevertheless expects that the ambient  humidity plays a role on the drying kinetic, as $a_e$ sets the limiting concentration $\varphi^\star$ at the drying interface.

However, Roger {\it et al.} investigated recently the drying of aqueous solutions of amphiphilic molecules
exhibiting self-assembled phases at high solute concentrations, for mimicking the biological complexity of water evaporation through mammalian skins~\cite{Spaar:00,Roger:16}. Strikingly, their experiments demonstrated that evaporation rates from such complex fluids do not depend on the humidity (at long time scales and on a wide range $a_e = 0$--0.95), as also observed for real mammalians skins. Their main interpretation, strengthened by in-situ observations, relies on the nucleation and growth of a thin concentrated phase at the drying interface, leading to the conclusion that self-assembly is possibly a key ingredient to explain this humidity-insensitive regime~\cite{Trabesinger:16}.

In the present work, we show actually that evaporation rates do not depend  on the ambient humidity $a_e$ (on a wide range) in the falling rate regime, 
for complex fluids such as solutions of polymers or surfactants, independently on any phase transition. Surprisingly, such a result has not been
demonstrated theoretically to the best of our knowledge, despite the large  amount of previously reported theoretical
investigations, probably because such works mainly focused on the description of thin film drying~\cite{Saure98,Guerrier:98,Okazaki:74}.
We demonstrate  in the following, using both numerical simulations of mass transport equations and analytical solutions in some limiting cases, that evaporation rates are nearly humidity-insensitive  when two ingredients are encountered: (1) a sharp
decrease of the water chemical activity at high solute concentration, and/or (2) a strong decrease of the mutual diffusion coefficient of the aqueous solution at high solute concentration. These features are commonly encountered for solutions of polymers and amphiphilic molecules thus emphasizing the significance of this result for processes involving water transport within complex fluids. 

\begin{figure}[ht]
\begin{center}
\includegraphics{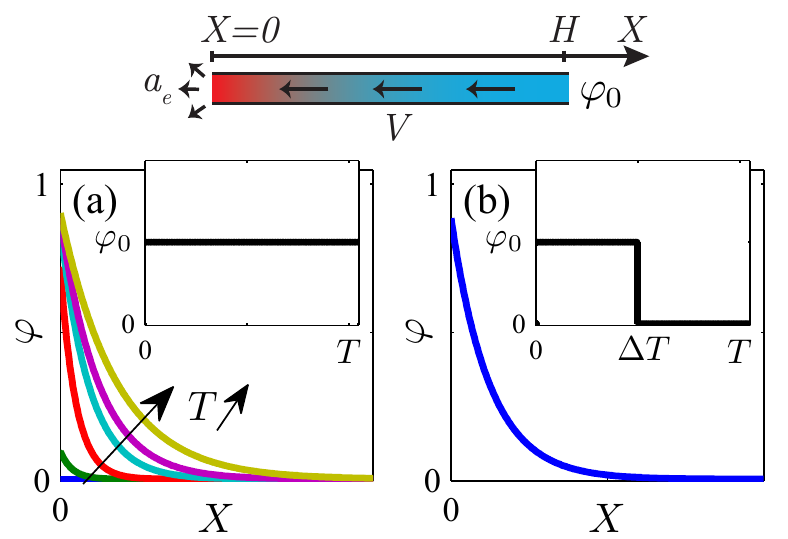}
\caption{Uni-directional drying of a binary mixture within a capillary (length $H$). Top: solvent evaporation at a rate $V_{ev}$ induces a flow which enriches the tip with solutes 
(colors code for the solute concentration). $\varphi_0$ is the concentration imposed at the inlet of the capillary, $a_e$ the ambient humidity. 
(a) and (b) Two different experimental configurations. (a) Growing concentration polarization layer, i.e. fixed concentration $\varphi_0$ in the reservoir: solutes accumulate continuously within the capillary (see inset).
(b) Steady concentration polarization layer obtained when imposing $\varphi_0 = 0$ in the reservoir after a delay time $\Delta T$, see inset.  
\label{fig:Sketch}}
\end{center}
\end{figure}

To demonstrate this result, we will consider for simplicity the case of a binary mixture composed of a volatile solvent and a non-volatile solute, and  we will focus more precisely on the model experiments shown 
schematically in Fig.~\ref{fig:Sketch}. The solution
is confined within a long capillary (length $H$) with an open end from which solvent evaporates, and is connected to a reservoir at the opposite end.
 We will also consider two different experimental configurations, see Fig.~\ref{fig:Sketch}(a) and (b). 
In the first case, the reservoir contains solutes at a volume fraction $\varphi_0 >0$, and evaporation continuously concentrates the non-volatile solutes up to the tip of the capillary. 
This geometry exactly  corresponds to the experiments reported in Ref.~\cite{Roger:16}. 
This case of {\it uni-directional} drying, is a common model for many other drying configurations: suspended or confined drops~\cite{Brutin:2015,Daubersies:11},  
pervaporation in microfluidic geometries~\cite{Daubersies:13}, as well as a strictly equivalent model of solvent evaporation from quiescent thick films with a moving interface  
in the regime $H \gg  D_0/V_{ev}$, see Fig.~\ref{fig:Sketch1}, Sec.~\ref{sec:theo} and appendix \ref{AppA}. 

In the second configuration, solutes contained within the reservoir at a concentration $\varphi_0 >0$ are replaced by pure water after a delay time $\Delta T$, i.e. $\varphi_0 = 0$
for $T>\Delta T$. The solutes previously trapped  within the capillary reach a steady concentration gradient, 
see Fig.~\ref{fig:Sketch}(b).
This steady configuration may be relevant for describing many different experimental cases, such as (i) the steady water transport through a polymer coating, an important issue in the field of waterproofing coatings for instance,
(ii) evaporation-induced steady water transport through hydrogel-based membranes \cite{Jeck:11}, or even (iii) through mammalian skins as suggested recently~\cite{Roger:16}. These examples may be also key to describe evaporation from soft contact lenses~\cite{Fornasiero:08},
or from any  other biological tissue exposed to air, such as eye's cornea for instance. These last examples may however involve contributions not taken into account in the model developed later (e.g. elastic effects~\cite{Okunozo:06,Okuzono:08}), and we will briefly discuss these issues in our conclusion as   
future research perspectives.

We will in the following investigate the role of the ambient humidity on the  steady evaporation rate, for a constant {\it volume} of solutes trapped within the capillary, defined as:
 \begin{eqnarray}
\Psi = \int_0^\infty  \varphi(X) \;  \text{d}X\,, \label{eq:psidim}
\end{eqnarray}  
where $\varphi(X)$ is the  steady solute volume fraction profile. The volume of solutes $\Psi$ trapped within the capillary depends directly on $\Delta T$ and 
on the initial concentration $\varphi_0$ during the {\it feeding} stage.
This second configuration corresponds to water transport driven by evaporation through a steady  concentration polarization layer, thus maintained by the competition between evaporation-induced convection and molecular diffusion.   Note also that the two configurations described above are fundamentally different, as no steady regime is reached in the first configuration, see Fig.~\ref{fig:Sketch}(a).

\section{Theoretical modeling of unidirectional drying \label{sec:theo}}

\subsection{Transport equations for  a binary mixture}

We consider a binary mixture composed of a volatile solvent and a non-volatile solute, and we define $V_1$ and $V_2$ the solute and the solvent velocities respectively. For simplicity, we also assume additivity of the volumes, i.e.
$1/\rho = w/\rho^0_1 + (1-w)/\rho^0_2$, where $\rho^0_i$ are the densities of the pure solute and solvent, $\rho$ the density of the mixture, and $w$ the solute mass fraction.
In the reference frame of the volume-averaged velocity defined as 
$\mathbf{V} = \varphi \mathbf{V}_1 + (1-\varphi)\mathbf{V}_2$, mass conservation equations for the solvent and the solutes are~\cite{Bird}:
 \begin{eqnarray}
&&\partial_T \varphi + \mathbf{V}.\nabla \varphi = \nabla (D(\varphi)\nabla\varphi)\,, \label{eq:gentrans}\\
&&\nabla . \mathbf{V} = 0\,, \label{eq:divV}
\end{eqnarray}
where $D(\varphi)$ is the mutual diffusion coefficient of the mixture (also called collective diffusion 
coefficient in the field of colloidal dispersions).

In the geometry displayed in Fig.~\ref{fig:Sketch}, we assume that the only flow within the capillary is due to water evaporation  (e.g. no buoyancy-driven flows, no Marangoni flows at the tip, etc.), and that concentrations are homogeneous across the transverse dimensions of the capillary, allowing us to reduce Eqs.~(\ref{eq:gentrans}) and (\ref{eq:divV}) to one
dimensional equations only. 
By convention, the evaporation rate $V_{ev}$ is positive for evaporation. 
With the axis defined in Fig.~\ref{fig:Sketch}, mass conservation (\ref{eq:divV}) 
results in a uniform flow of velocity:
\begin{equation}
V=-V_{ev}\,,
\end{equation}
within the capillary, from the reservoir up to its tip.
We also assume, as done classically, see also above Eq.~(\ref{eva2}), that $V_{ev}$ is given by: 
 \begin{eqnarray}
V _{ev} = (a(\varphi_i)-a_e)\,J\,, \label{eq:aphi}
\end{eqnarray}	
where $a(\varphi)$ is the solvent chemical activity, $a_e$ the ambient  humidity (for non aqueous solvents, $a_e = 0$ {\it a priori}), and $\varphi_i = \varphi(X=0)$ is the concentration at the interface.
$J>0$ (m/s) is a coefficient which depends on mass transfer in the gas phase~\cite{Cussler}, and it corresponds to the evaporation-induced flow in the case of pure water and for $a_e=0$. 
The term $a(\varphi_i)-a_e$ accounts for the diminution of the evaporation driving force due to the decrease of the solvent chemical 
activity at the interface. This evaporation-driven flow  
convects in turn the solute contained in the reservoir up to the channel tip where they accumulate. 
The solute concentration profile within the channel follows Eq.~(\ref{eq:gentrans}) which reduces to the 1D equation:
 \begin{eqnarray}
\partial_T \varphi + V\partial_X \varphi = \partial_X (D(\varphi)\partial_X\varphi)\,, \label{eq:trans1Ddim}
\end{eqnarray}
where $V$ is the drying-induced flow (m/s) within the capillary.

In the above equations, we have assumed implicitly {\it local thermodynamic equilibrium} conditions, 
thus discarding evaporation-induced glass transition as observed for some polymeric systems~\cite{Degennes:2002,Okunozo:06,Okuzono:08}, and other possible kinetic effects such as the nucleation and growth of crystallites.  However, the model given by Eqs.~(\ref{eq:aphi}-\ref{eq:trans1Ddim}) can be used to  describe binary systems exhibiting self-assembled phases, still within the assumption of local thermodynamic equilibrium, providing that proper continuity conditions are applied at the phase boundaries, see for instance Ref.~\cite{Schindler:09} in a similar context.

\subsection{Dimensionless model}
We first define $D(\varphi) = D_0 \hat{D}(\varphi)$ with $\hat{D}(\varphi\to 0)\to 1$.
As mentioned above, our analysis focuses on the regime $H \gg  D_0/V_{ev}$, and the geometry can be described by a semi-infinite medium. Therefore, we did not use the length $H$ to make equations dimensionless, but rather the relevant scale $D_0/J$ to define the following unitless dimensions:
\begin{eqnarray}
&&x = X/(D_0/J)~~\text{and}~~~t = T/(D_0/J^2)\,. \label{eq:Varadim}
\end{eqnarray}
The above 1D transport equations become:
 \begin{eqnarray}
&&\partial_t \varphi = v \partial_x \varphi + \partial_x (\hat{D}(\varphi)\partial_x\varphi)\,, \label{eq:trans1}\\ 
&&v = a(\varphi_i)-a_e \label{eq:trans2}\,,
\end{eqnarray}
where $v=V_{ev}/J = -V/J$. 
The non-volatility of the solute imposes the following boundary condition at the air/solution interface: 
\begin{eqnarray}
\varphi_i\,v  + \hat D(\varphi_i) (\partial_x \varphi)_{x=0} = 0\,, \label{eq:BC1} 
\end{eqnarray}
and we assume a semi-infinite medium:
 \begin{eqnarray}
\varphi(x\to \infty, t) = \varphi_0\,, \label{eq:BC2}
\end{eqnarray}
corresponding to a fixed concentration at the opposite inlet of the capillary at any time $t$. 
The initial condition is:
 \begin{eqnarray}
\varphi(x, t=0) = \varphi_0\,. \label{eq:IC}
\end{eqnarray}

\subsection{Empirical laws for $a(\varphi)$ and $D(\varphi)$\label{sec:EmpLaws}}
The drying kinetics a priori depends on thermodynamic parameters ($a(\varphi)$, $a_e$), on mass transport within the liquid
phase ($D(\varphi)$), but also on the transport in the vapor phase ($J$). 
We consider, in the following, binary mixtures for which 
 $a(\varphi\to 1) \to 0$  and $D(\varphi)$  decreasing (possibly strongly) for $\varphi \to 1$. These features are indeed common for a wide range of experimental binary mixtures, 
including particularly solutions of polymers, copolymers and amphiphilic molecules~\cite{neogi96,Flory,Degennes,Daubersies:11,Daubersies:13,GU2005,Doumenc:05,Okazaki:74,Jeck:11}. As shown later, our results do not depend strongly on the
exact shapes of $a(\varphi)$ and $D(\varphi)$, and we will thus take empirical formulae for a given system to capture
the key features of the drying dynamics for most of binary molecular complex fluids. 

More precisely, we compiled different sets of measurements 
of both $a(\varphi)$ and $D(\varphi)$ for poly(vinyl
alcohol) (PVA) aqueous solutions~\cite{Okazaki:74,Jeck:11}.
These precise measurements obtained at $25^\circ$C using different techniques (time-resolved measurements of sorption kinetics, micro-interferometric methods) and for different commercial grades, 
lead to very close data sets which are well-fitted by the following relations:
\begin{eqnarray}
\log_{10} D(\varphi) &=& a_4\varphi^4+a_3\varphi^3+a_2\varphi^2+a_1\varphi+a_0, \label{eq:D}
\end{eqnarray}
with $[a_4;a_3;a_2;a_1;a_0]=[-13.65; 17.47; -8.97; 1.12; -10.29]$,
and
\begin{eqnarray}
&&a(\varphi)  =(1-\varphi)\exp(\varphi + \chi \varphi^2)\,, \label{actPVA}\\
&&\chi(\varphi) = 3.94 - 3.42(1-\varphi)^{0.09}\,. \label{actPVA2}
\end{eqnarray}
The latter relation corresponds to the solvent activity in a polymer solution within the framework of the Flory-Huggins theory, and
$\chi(\varphi)$ is the binary Flory-Huggins interaction parameter~\cite{Flory}.

\begin{figure}[ht]
\begin{center}
\includegraphics{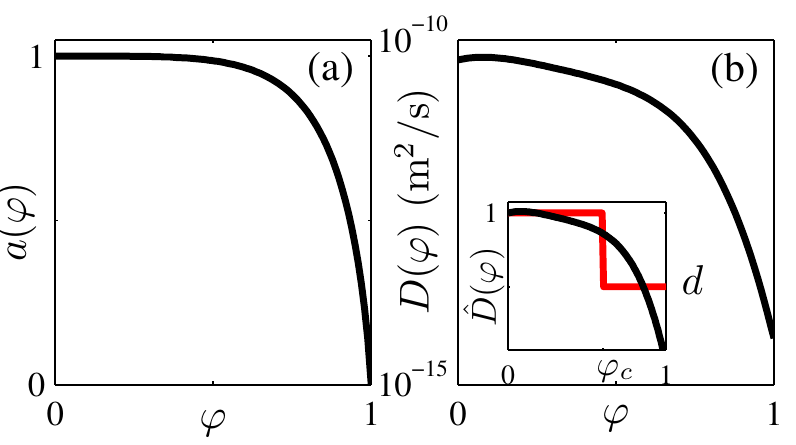}
\caption{Properties of the binary system 
PVA/water. (a) Chemical water activity and (b) mutual diffusion coefficient, see Eqs.~(\ref{eq:D}--\ref{actPVA2}) and Refs.~\cite{Okazaki:74,Jeck:11} for the corresponding measurements. 
Inset of (b): normalized mutual diffusion coefficient $\hat{D}(\varphi)$ (black line), and piecewise constant  $\hat{D}(\varphi)$ (red line, see text).
\label{fig:Da}}
\end{center}
\end{figure}

These two relations are displayed in Fig.~\ref{fig:Da}, and we will use them as representative empirical laws for many other 
complex fluids.

\section{Steady polarization layer} \label{SteadyCase}

We first consider for simplicity the configuration shown in Fig.~\ref{fig:Sketch}(b).
It corresponds to a steady situation obtained when the concentration $\varphi_0$ at the inlet has been set at $\varphi_0 = 0$  after a delay time, see the inset of Fig.~\ref{fig:Sketch}(b). 
After a transient, solutes reach a steady  concentration
profile where solute convection exactly balances diffusion.
From Eqs.~(\ref{eq:trans1},\ref{eq:BC1}), the concentration profile then follows:
 \begin{eqnarray}
&&0 = v \varphi + \hat{D}(\varphi)\partial_x\varphi \label{eq:steady1} \, ,
\end{eqnarray}
where $v$ is given by Eq.~(\ref{eq:trans2}).
We will investigate in the next section the dependence of the drying rate $v$ along with the humidity $a_e$, while keeping constant the volume of solute $\Psi$
trapped within the capillary, see Eq.~(\ref{eq:psidim}). We further define the unitless volume of solutes as:
 \begin{eqnarray}
\psi = \int_0^\infty  \varphi(x) \;  \text{d}x\,,\label{eq:psi}
\end{eqnarray}  
and we will consider constant $\psi$ values in the following.
As shown below, the volume of solutes $\psi$ governs both the evaporation rate and  the extent of the polarization layer. The  assumption of a semi-infinite medium, i.e. $H \gg  D_0/V_{ev}$ with real units, thus sets a maximal value for $\psi$ which depends on $H$.

\subsection{General case: non-constant $a(\varphi)$ and  non-constant $\hat{D}(\varphi)$} \label{SteadyGen}
We first consider the most general case corresponding  to non-constant $a(\varphi)$ and $D(\varphi)$, and given by the empirical formulae (\ref{eq:D}--\ref{actPVA2}) displayed in Fig.~\ref{fig:Da}. 
We solved the non-linear equation~Eq.~(\ref{eq:steady1}) with $v=a(\varphi_i)-a_e$ for several  humidities $a_e$, and for several $\psi$ values, see appendix \ref{AppRes} for  details.
\begin{figure}[ht]
\begin{center}
\includegraphics{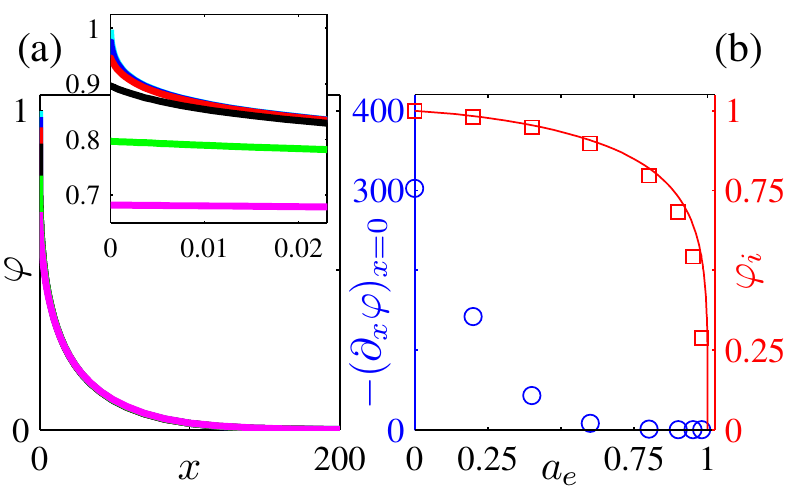}
\caption{Steady polarization layer: non-constant $a(\varphi)$ and  non-constant $\hat{D}(\varphi)$. (a) Steady profiles $\varphi(x)$ for $\psi = 15$ and several ambient humidity $a_e$ 
(0: cyan; 0.2: blue; 0.4: red; 0.6: black; 0.8: green; 0.9: magenta).  Insert: same data but at low $x$. (b) 
Concentration gradient at the interface $-(\partial_x\varphi(x))_{x=0}$ vs. $a_e$ ($\circ$, left axis), and concentration at the interface $\varphi_i$ vs. $a_e$ ($\square$, right axis), for $\psi = 15$. 
The solid line is the theoretical curve $\varphi^\star$ vs. $a_e$ with $a(\varphi^\star) = a_e$. 
} 
\label{fig:Steady}
\end{center}
\end{figure}
We will consider $\psi \gg 1$ leading to evaporation rates $v \ll 1$, owing to the significant decrease of the chemical activity at the drying interface.
As shown in Fig.~\ref{fig:Steady}(a) for the case $\psi = 15$, the profiles almost collapse on a 
single curve for ambient humidities $a_e\leq 0.9$ except in a narrow region of small $x$ as shown in the inset.
The data shown in Fig.~\ref{fig:Steady}(b), $-(\partial_x \varphi)_{x=0}$ vs. $a_e$,  help to reveal the strong concentration gradient at the interface for  $a_e \to 0$ (over 4 decades for $a_e$ ranging from 0.95 to 0). 
Fig.~\ref{fig:Steady}(b) also shows that the concentration at the interface $\varphi_i$ is close to $\varphi^\star$ given by $a(\varphi^\star) = a_e$, indicating thus that the evaporation rates are indeed small for such large $\psi$ value.

\begin{figure}[ht]
\begin{center}
\includegraphics{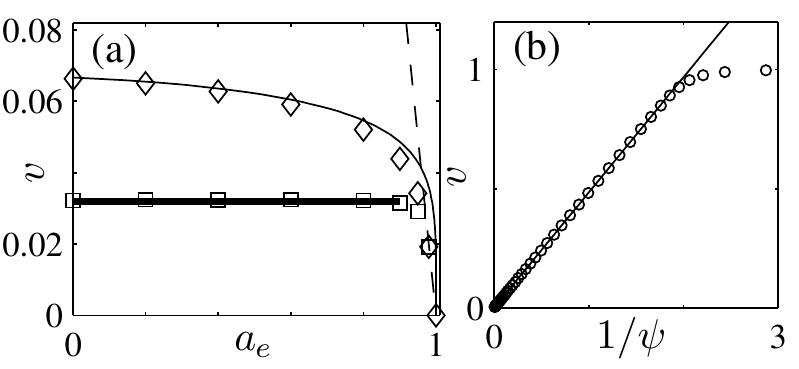}
\caption{(a) Steady evaporation rate $v$ vs. $a_e$. $\square$~: case $\hat{D}(\varphi)$ given by Eq.~(\ref{eq:D}) for $\psi = 15$, see also 
the corresponding profiles in Fig.~\ref{fig:Steady}.
The thick dark line is $v = \varphi_c / \psi$ with $\varphi_c = 0.48$.
$\diamond$:~case $\hat{D}(\varphi) = 1$ for $\psi = 15$.  The solid line is the theoretical prediction $v = \varphi^\star/\psi$ with $a(\varphi^\star) = a_e$. 
The dashed line is the case expected for a pure solvent i.e. $v = 1-a_e$. 
(b) Steady evaporation rate $v$ at $a_e=0$ vs. $1/\psi$ for the case $\hat{D}(\varphi)$ given by Eq.~(\ref{eq:D}). The solid line is the best 
fit by $v = \varphi_c / \psi$ with  $\varphi_c = 0.48$.}
\label{fig:Case2states}
\end{center}
\end{figure}

The corresponding evaporation rates $v$ are shown against the ambient humidity $a_e$ in Fig.~\ref{fig:Case2states}(a) for $\psi = 15$. 
As pointed out in the introduction, the evaporation rates are nearly constant over a wide range of $a_e$  (relative decrease $\simeq 2\%$ for $a_e=0$--0.9, $\psi = 15$).  
Fig.~\ref{fig:Case2states}(b) also shows the calculated evaporation rates $v$ for $a_e=0$, vs. $1/\psi$. These data show that for $\psi > 1$, 
evaporation rates are very well-fitted by $v \simeq \varphi_c / \psi$ with $\varphi_c = 0.48$, suggesting possibly that simple analytical expressions can be found to explain why evaporation rates are insensitive to $a_e$. 

With real units, the evaporation rate follows $V_{ev} \simeq D_0 \varphi_c /\Psi$ independently of the 
humidity ($\Psi$ being the dimensionalized  volume of solutes trapped in the capillary) and of the transport in the vapor phase (term $J$). 
Water evaporation however still depends on diffusion in the liquid phase (term $D_0$) and as shown later, the exact value $\varphi_c$ accounts for the specific shape of $D(\varphi)$.

We unveil below the role played by both  $\hat{D}(\varphi)$ and $a(\varphi)$ on 
the very small dependence of $v$ with $a_e$,  
using analytical solutions obtained for simple expressions of $\hat{D}(\varphi)$.

\subsection{Role of the activity only: case $\hat{D}(\varphi) = 1$} \label{SteadyActi}

In this case, concentration profiles are easily calculated from Eq.~(\ref{eq:steady1}) with the constraint given by Eq.~(\ref{eq:psi}), leading to the exponential decay:
\begin{eqnarray}
	\varphi(x) = \psi v \exp(-v x)\,.  \label{ExpDecay}
\end{eqnarray} 
For $\psi \gg 1$, one should have $v\ll 1$ to get a finite concentration at the interface, and thus $\varphi_i = \psi v \simeq \varphi^\star$ with $\varphi^\star$ given by $a(\varphi^\star) = a_e$. In that regime, the evaporation rate thus follows 
\begin{equation}
	v \simeq \frac{\varphi^\star}{\psi}\,, \label{RateDconst}
\end{equation}
($V_{ev} \simeq D_0 \varphi^\star /\Psi$ with real units). The  water/PVA case is shown in Fig.~\ref{fig:Case2states}(a) for $\psi = 15$ and assuming
$\hat{D}(\varphi) = 1$. Symbols  are the direct solutions of $\psi v = \varphi_i$ with Eq.~(\ref{eq:trans2}) and the solid line
is the theoretical approximation (\ref{RateDconst}).  
With the assumption of a constant mutual diffusion coefficient, evaporation rates are weakly sensitive 
to the humidity variations (relative variations of $<20\%$ over $a_e=0$--0.8 for $\psi = 15$). 

Indeed, a sharp decrease of the chemical activity at high solute concentrations (see Fig.~\ref{fig:Da}(a)) 
leads to $\varphi^\star \simeq 1$ over a large humidity range (see Fig.~\ref{fig:Steady}(b)). 
A weak dependance of the evaporation rate follows, because of Eq.~(\ref{RateDconst}). 
Nevertheless, the results reported in Fig.~\ref{fig:Case2states} show that the variations of $D$ with $\varphi$ are key to understand the extremely small variations 
of the evaporation rate along with $a_e$.  
  
\subsection{Role of the mutual diffusion coefficient: non-constant $a(\varphi)$ and piecewise constant  $\hat{D}(\varphi)$}  \label{SteadyD}

To capture the role played by the variation of the mutual diffusion coefficient with the solute concentration, we solve analytically 
the steady case Eq.~(\ref{eq:steady1}) using a piecewise constant function for $\hat{D}(\varphi)$. More precisely, we chose:
\begin{equation}
\left . 
	\begin{aligned}
&&\hat{D}(\varphi) = 1~~\text{for}~~\varphi<\varphi_c \\
&&\hat{D}(\varphi) = d~~\text{for}~~\varphi>\varphi_c 
	\end{aligned}
\right \}  
\label{eq:piecewise}
\end{equation} 
with $d<1$, as shown in the inset of Fig.~\ref{fig:Da}(b). We also consider humidities $a_e<a(\varphi_c)$ and large $\psi$ values for which the concentration at the drying interface reaches $\varphi_i>\varphi_c$.
Integration
of Eq.~(\ref{eq:steady1}) in this simple configuration leads to the following concentration profile:
\begin{eqnarray}
	&&\varphi(x) = \varphi_i \exp(-v x /d)~\text{for}~x<x_c\,,  \label{Sol1} \\	
	&&\varphi(x) = \varphi_c \exp(-v (x-x_c))~\text{for}~x>x_c\,. 		\label{Sol2}
\end{eqnarray}
The concentration field continuity imposes $x_c = (d/v)\log(\varphi_i/\varphi_c)$.
Integrating Eqs.~(\ref{Sol1}-\ref{Sol2}) from $x=0$ to infinity and using the definition of $\psi$ (Eq.~(\ref{eq:psi})) yields: 
\begin{eqnarray}
	v = \frac{\varphi_c + d(\varphi_i - \varphi_c)}{\psi}\,. \label{eq:vdphic}
\end{eqnarray}
In the general case, $\varphi_i$ and $v$ are obtained by solving numerically the set of algebraic equations (\ref{eq:trans2},\ref{eq:vdphic}).
For large $\psi$, Eq.~(\ref{eq:trans2}) can be replaced by $\varphi_i  \simeq \varphi^\star$, 
and Eq.~(\ref{eq:vdphic}) directly gives the evaporation rate $v$.
As expected, Eq.~(\ref{RateDconst}) is recovered for $d=1$. 
For $d \ll 1$, Eq.~(\ref{eq:vdphic}) reduces to:
\begin{eqnarray}
	v \simeq \frac{\varphi_c}{\psi}\,, \label{VevDconst}
\end{eqnarray}
where $v$ no longer depends on $\varphi^\star$.
We show in the following that this crude model based on a piecewise constant $D(\varphi)$ can be used to explain most of the features of more general  cases.

The case $d \to 0$ corresponds to the existence close to the drying interface of a layer of vanishing thickness $x_c \simeq (d/v)\log(\varphi^\star/\varphi_c)$
with a diverging concentration gradient at the interface: $(\partial_x \varphi)_{x=0} \simeq - \varphi^\star v /d$, from Eq.~(\ref{Sol1}). 
For larger $x$ values, i.e. $x>x_c$, concentration profiles collapse on a master curve, whatever the value of the humidity, with a smaller concentration gradient on a scale $1/v \gg x_c$, see Eq.~(\ref{Sol2}).
These behaviors were also observed for the PVA case studied in section \ref{SteadyGen} (Fig.~\ref{fig:Steady}). 
Furthermore, an equivalence between the simplified model and this more realistic case can be found by fitting $\varphi_c$ in Eq.~(\ref{VevDconst}) from the curve $v$ vs. $1/\psi$ of Fig.~\ref{fig:Case2states}(b)
(PVA case with $a_e=0$). 
We get $\varphi_c \simeq 0.48$.
For the PVA case, this specific value corresponds approximately to the concentration below which all the
profiles collapse at large $x$, whatever the value of the humidity $a_e$, see Fig.~\ref{fig:Steady}(a).


We can now draw a simple picture to explain why evaporation rates are nearly insensitive to variations of $a_e$ in this steady configuration.
Concentrations at the interface reach $\varphi_i \simeq \varphi^\star$ for $\psi \gg 1$, and the profiles display strong concentration gradients at the interface,
owing to the very small mutual diffusion coefficient. 
Concentrations thus drop from $\varphi^\star$ to values $\varphi \leq \varphi_c$ on a very thin layer (0--$x_c$), and the contribution of this strong gradient to the total amount of solutes trapped within the capillary is negligible, see Eq.~(\ref{eq:psi}). 
The value of the ambient humidity only plays a role on this thin layer through $\varphi^\star$ and $x_c$. 
For the remaining part of the profile (i.e. $x\geq x_c$), mutual diffusion coefficient is almost constant ($D_0$ with real units in the simple model). 
Concentration profiles in this region are solutions of a diffusion problem with constant diffusivity and concentration $\varphi_c$ imposed at $x=x_c$.
When $x_c$ is very small, this is almost equivalent to imposing $\varphi_c$ at $x=0$.
The effect of humidity then drops out, because $\varphi_c$ is a material property, hence independent of ambient conditions.

Practically, vanishing diffusivity ($d \to 0$) is not strictly required to get evaporation rates nearly insensitive to ambient humidity.  
Indeed, assuming for instance $d = 0.1$, $\varphi_c = 0.5$  and $\varphi_i \simeq \varphi^\star$ (as
expected for large $\psi$) in Eq.~(\ref{eq:vdphic}), $v$ only decreases of $\simeq 3\%$ over $a_e = 0$--0.8. 
This weak variation comes indeed from two superimposed effects: 
(i) the small variation of the concentration at the interface $\varphi_i$ along with the humidity (due to the sharp variation of $a(\varphi)$ vs. $\varphi$ at high solute concentration), 
and (ii) a small mutual diffusion coefficient at high solute concentration 
(presence of a thin layer with a high concentration gradient at the drying interface).

\section{Growing polarization layer ($\varphi_0 > 0$)}

We now turn to the time-dependent case described schematically in Fig.~\ref{fig:Sketch}(a). In this regime,
concentration in the reservoir is $\varphi_0 \ll 1$, and solutes continuously accumulate within the capillary.
At early stages, the drying-induced flow convects solutes at the tip of the capillary where they accumulate in a region of size $\xi \sim 1$ ($\xi \sim D_0/V_{ev}$ with real units).  
At longer time scales however, concentration at the tip is expected to increase asymptotically towards $\varphi^\star$, given by $a(\varphi^\star) = a_e$, 
leading to a slowing down of the evaporation rate (because of Eq.~(\ref{eq:trans2})),
and thus of the convective flux $-(\varphi v)$. 
One thus expects a broadening of the concentrated layer as the latter evolves as $\xi \sim D_0/V_{ev}$, and we  
will mainly focus in the following on this long-time scales regime.
Note again that the assumption of a semi-infinite medium, i.e. $H \gg  D_0/V_{ev}$ see above, imposes a limiting time scales for our description which depends on $H$,
as the evaporation rate $V_{ev}$ is expected to slow down continuously at long time scales. 
Note also that this time-dependent case is strictly analogous to the drying of quiescent thick films, as for instance studied in Ref.~\cite{doumenc04}. 
One can indeed show using an appropriate change of variables, 
that Eqs.~(\ref{eq:trans1}--\ref{eq:trans2}) along with the boundary and initial conditions (\ref{eq:BC1}--\ref{eq:IC}) corresponds to the modeling of solvent evaporation from a thick film, see appendix \ref{AppA}.

\subsection{General case: non-constant $a(\varphi)$ and  non-constant $\hat{D}(\varphi)$}

\begin{figure}[ht]
\begin{center}
\includegraphics{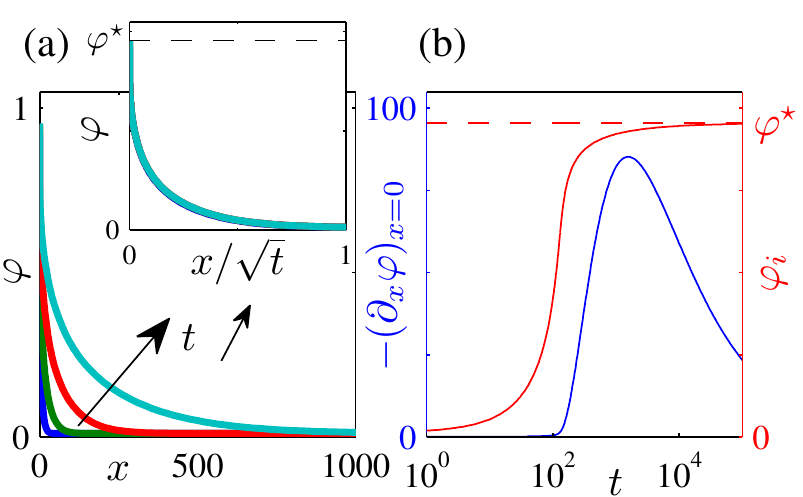}
\caption{Growing polarization layer: non-constant $a(\varphi)$ and  non-constant $\hat{D}(\varphi)$. (a): $\varphi(x)$ vs. $x$ for several times $t=10^3$; $10^4$, $10^5$
 and $10^6$ ($a_e= 0.4$, $\varphi_0 = 0.01$) for the water/PVA case.
The inset displays the same data rescaled against $x/\sqrt{t}$, $\varphi^\star$ is given by $a(\varphi^\star) = 0.4$. 
(b) Concentration gradient at the interface $-(\partial_x\varphi(x))_{x=0}$ vs. $t$ (left axis), and concentration at the interface $\varphi_i$ vs. $t$ (right axis)}  
\label{fig:CasePVA1}
\end{center}
\end{figure}
We used a commercial finite elements software, Comsol Multiphysics, to solve numerically the transport equations~(\ref{eq:trans1}--\ref{eq:trans2}) with the boundary and initial conditions~(\ref{eq:BC1}--\ref{eq:IC}), 
assuming the empirical formulae (\ref{eq:D}--\ref{actPVA2}) for $a(\varphi)$ and $\hat{D}(\varphi)$ 
(see Fig.~\ref{fig:Da}). 
Comsol Multiphysics is based on the Galerkin method. We used quadratic Lagrange elements and BDF solver.

Figure~\ref{fig:CasePVA1} displays some concentration profiles computed for $a_e = 0.4$ and $\varphi_0 = 0.01$ at long time scales ($t>1000$). In this time-dependant case, concentration reaches 
$\varphi_i \simeq \varphi^\star$ given by $a(\varphi^\star) = a_e$ for $t>1000$. 
Concentration profiles then widen over time, and again with steep concentration gradients at the interface (which ultimately decrease owing to the 
decrease of the evaporation rate).
Concentration profiles almost collapse on a single curve when plotted against $x/\sqrt{t}$ suggesting that a self-similar profile approximates the calculated solutions in this long-time scale regime.

\begin{figure}[ht]
\begin{center}
\includegraphics{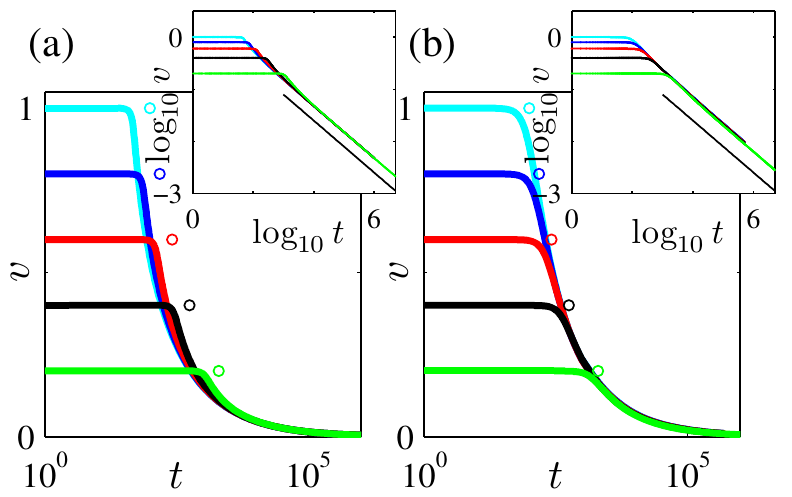}
\caption{Evaporation rate $v$ vs. $t$ for several ambient humidities $a_e$ (0: cyan; 0.2: blue; 0.4: red; 0.6: black; 0.8: green). (a)  water/PVA general case, (b) case $\hat{D}(\varphi) =1$.
The circles are simple estimates of the onset of the falling rate period
assuming constant $D(\varphi)$ and $v=1-a_e$ at early time scales, see text and appendix~\ref{SoluceApp} for details. 
The insets display the same data in a log-log plot. The black lines show the behaviors $v\sim 1/\sqrt{t}$.} 
\label{fig:CasePVA2}
\end{center}
\end{figure}

The corresponding temporal evolution of the evaporation rate $v$ is shown in Fig.~\ref{fig:CasePVA2}(a) for various humidities. 
 At early stages, $v \simeq 1 - a_e$ as the solute concentration at the interface does not modify strongly  the water chemical activity (the so-called {\it constant rate regime}). At longer time scales, evaporation rates collapse on a master curve regardless of the value of the humidity.  In this asymptotic regime, the log-log plot in Fig.~\ref{fig:CasePVA2}(a)
helps to reveal that the evaporation rate decreases as 
\begin{eqnarray}
	v \simeq \frac{\alpha}{\sqrt{t}}\,, \label{fitalpha}
\end{eqnarray}
 as also reported by Roger {\it et al.} for water/surfactant  
solutions, in an experimental configuration close to the one studied here~\cite{Roger:16}.

\begin{figure}[ht]
\begin{center}
\includegraphics{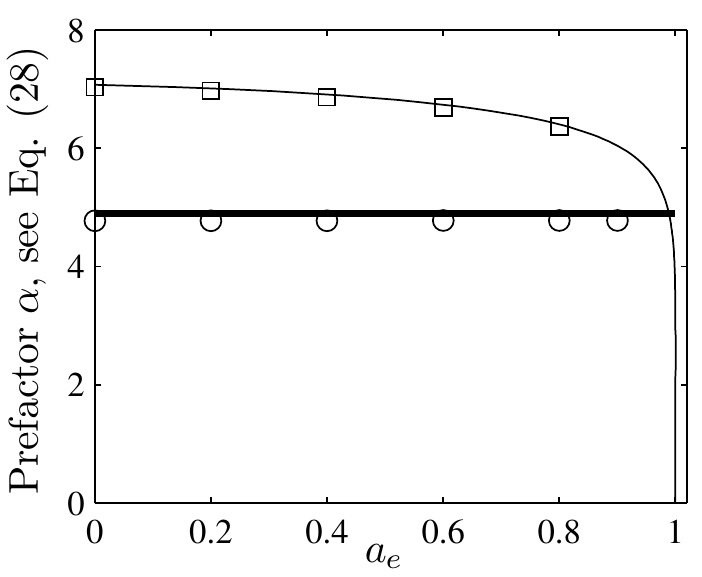}
\caption{Prefactor $\alpha$ vs. ambient humidities $a_e$. $\alpha$ is estimated from the evaporation rate in the asymptotic regime using Eq.~(\ref{fitalpha}), i.e. $v \simeq \alpha/\sqrt{t}$;  $\circ$: water/PVA case, $\square$: case $\hat{D}(\varphi)=1$.
The thin line is the theoretical prediction Eq.~(\ref{vselsimilar}). The thick line is the prediction given by Eq.~(\ref{predictAlpha}) with $\varphi_c \simeq 0.48$.} 
\label{fig:CaseConstantDevap}
\end{center}
\end{figure}
Figure~\ref{fig:CaseConstantDevap} displays the prefactor $\alpha$ fitted from the data $v$ vs. $t$ at long time scales using Eq.~(\ref{fitalpha}), and against $a_e$.
$\alpha$ is extremely insensitive to the variations of $a_e$ (variations below $<0.1\%$ over the range $a_e = 0$--0.9) suggesting again that humidity does not play any role (at long time scales) on the evaporation kinetics, even in this time-dependent configuration.
To unveil again the role played by both the shape of the solvent chemical activity $a(\varphi)$ and that of the 
mutual diffusion coefficient $\hat{D}(\varphi)$, we turn to simple expressions of  $\hat{D}(\varphi)$.

\subsection{Role of the activity only: case $\hat{D}(\varphi) = 1$}
To emphasize the role played by the activity only, we solved Eqs.~(\ref{eq:trans1}--\ref{eq:trans2}) with the boundary and initial conditions~(\ref{eq:BC1}--\ref{eq:IC}), but assuming $\hat{D}(\varphi) = 1$.

\begin{figure}[ht]
\begin{center}
\includegraphics{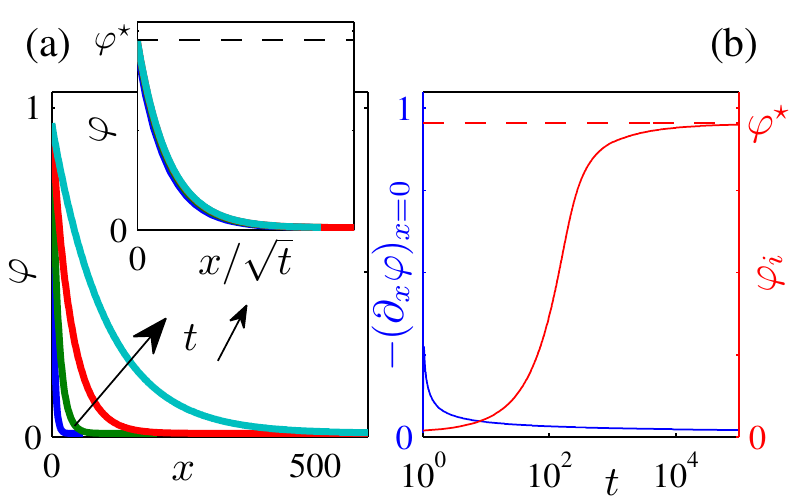}
\caption{Growing polarization layer: non-constant $a(\varphi)$ and  $\hat{D}(\varphi) = 1$.
(a): $\varphi(x)$ vs. $x$ for several times $t=10^3$; $7.9\times 10^3$, $6.3 \times  10^4$
 and $5\times 10^5$ ($a_e= 0.4$, $\varphi_0 = 0.01$.
The inset displays the same data rescaled against $x/\sqrt{t}$, $\varphi^\star$ is given by $a(\varphi^\star) = 0.4$. 
(b) Concentration gradient at the interface $-(\partial_x\varphi(x))_{x=0}$ vs. $t$ (left axis), and concentration at the interface $\varphi_i$ vs. $t$ (right axis)
}  
\label{fig:CaseContantD}
\end{center}
\end{figure}
Figure~\ref{fig:CaseContantD} shows some concentration profiles $\varphi(x)$ for $a_e = 0.4$ and $\varphi_0 = 0.01$
at different times.
As for the previous case, the concentration at the tip increases asymptotically towards $\varphi^\star$ given by 
$a(\varphi^\star)= a_e$, and the solute gradient then widens over time. 
However, these profiles differ significantly from those reported in Fig.~\ref{fig:CasePVA1}, as the concentration gradients at $x=0$ remain moderate.
For $t>1000$, concentration profiles almost collapse over the whole $x$ range onto a single curve when plotted against $x/\sqrt{t}$.

Figure~\ref{fig:CasePVA2}(b) shows the calculated evaporation rates for different ambient humidities $a_e$. At long time scales, all the dynamics  seem to collapse on a single curve in this log-log plot.
Again, the evaporation rate decreases over time and follows Eq.~(\ref{fitalpha}) apparently regardless   
of the humidity. Figure~\ref{fig:CaseConstantDevap} however reports the fitted $\alpha$ values against the humidity $a_e$. 
$\alpha$ decreases slightly of $\simeq 10\%$ over the  range $a_e = 0$--0.8.

In appendix~\ref{SoluceApp}, we provide analytical estimates of the onset $t_c$ of 
the falling rate period using simple arguments assuming $\hat{D}(\varphi) = 1$. These 
calculations yield
\begin{eqnarray}
t_c = \frac{\varphi^\star - \varphi_0}{(1-a_e)^2 \varphi_0}\,,
\end{eqnarray}
plotted together with the evaporation rates as a function of time in Fig.~\ref{fig:CasePVA2}(a) and (b) (circles).
This comparison helps us to reveal that these simple calculations yield correct estimates of the transition between the constant and the falling rate 
periods in the case $\hat{D}(\varphi) = 1$, but that the decrease of the evaporation rate
occurs slightly earlier in the case of a non-constant $\hat{D}(\varphi)$. This suggests that 
the mutual diffusion coefficient plays a role in shifting slightly the onset of the falling rate period.

To account for the results displayed in Fig.~\ref{fig:CaseConstantDevap}, we look for an analytical solution in this long-time scales regime, 
after replacing the boundary condition (\ref{eq:trans2}) by $\varphi_i = \varphi^\star$, as already done in section \ref{SteadyActi} for the steady problem. 
The derivation of a self-similar solution based on the variable $u(x,t)=x/2\sqrt{t}$ is detailed 
in appendix \ref{AppB}.
Equation~(\ref{soluce2}) gives the concentration profiles, classically expressed with erf functions.
With the approximation $\varphi_0 \ll 1$, the evaporation rate reads
\begin{eqnarray}
	v \simeq \frac{\alpha}{\sqrt t} = \sqrt{\frac{\varphi^\star}{2\varphi_0 t}}\,. \label{vselsimilar}
\end{eqnarray}
which fits correctly the values from the numerical solutions (see the thin line in Fig.~\ref{fig:CaseConstantDevap}).

Remarkably enough, applying a first order series expansion to the exact solution (\ref{soluce2}) at $u(x,t) \to 0$ yields:
\begin{eqnarray}
\varphi(x,t) = \varphi_0 + (\varphi^\star-\varphi_0) \exp \left [-v(t) \, x \right ] \, ,\label{soluce_app}
\end{eqnarray}
which is very similar to the steady problem solution (\ref{ExpDecay}).
This shows that the humidity insensitive regime is mainly driven by a balance between advection and diffusion (RHS terms of Eq.~(\ref{eq:trans1})), as for the steady polarization layer, see Sec.~\ref{SteadyCase}.

As for the steady polarization layer (Sec.~\ref{SteadyCase}), the small decrease of the evaporation kinetics with the increasing humidity
($\alpha$ term in Eq.~(\ref{vselsimilar})) is only due to the abrupt variation of the chemical activity at high solute concentration.
Note also that the sensitivity to humidity is weaker in the time-dependent case as $v \propto \sqrt{\varphi^\star}$, whereas
$v \propto \varphi^\star$ in the steady case. This leads for instance to variations of $\alpha$ of $\simeq 10\%$ only 
over the humidity range $a_e = 0$--0.8, without any variation of the mutual diffusion coefficient. Note again that sharper 
decrease of the chemical activity with $\varphi$ would have thus produced more constant evaporation kinetics regardless of $a_e$.
Nevertheless, the results displayed in Fig.~\ref{fig:CaseConstantDevap} suggest again that the decrease of the mutual diffusion coefficient at high concentration plays a major role to explain why the evaporation dynamics are almost independent of $a_e$.

\subsection{Role of the mutual diffusion coefficient: non-constant $a(\varphi)$ and piecewise constant  $\hat{D}(\varphi)$}
We looked for a self-similar solution of Eqs.~(\ref{eq:trans1},\ref{eq:BC1}--\ref{eq:IC}) with the simplified model of $\hat{D}(\varphi)$ (\ref{eq:piecewise}) and the boundary condition $\varphi_i = \varphi^\star$.
The derivation is detailed in appendix \ref{AppB}.
With the assumptions $\varphi_0 \ll 1$ and $d \ll 1$, the evaporation rate reads:
\begin{align} 
	v \simeq \sqrt{ \frac{\varphi_c}{2 \varphi_0 t }}\,.  \label{predictAlpha}
\end{align}
This simple prediction looks like Eq.~(\ref{vselsimilar}) with the substitution $\varphi^\star \to \varphi_c$.  
Figure~\ref{fig:CaseConstantDevap} compares this theoretical prediction with the numerical results, using the value $\varphi_c \simeq 0.48$ derived from the steady polarization layer investigated in Sec.~\ref{SteadyD}, and demonstrates
that the agreement is very good. 

The evaporation rate Eq.~(\ref{predictAlpha}) corresponds to the solution of a diffusion problem with a constant diffusivity and the concentration $\varphi_c$ imposed at the drying interface.
As for the steady polarization layer, the major part of the liquid phase is shielded from the gas phase by a thin layer of liquid with a large concentration gradient.
We therefore conclude that the same mechanisms are acting in both steady and growing polarization layer.

\section{Conclusions and discussions}

In the present paper, we have investigated theoretically the one-dimensional water transport induced by evaporation from a molecular mixture. At long time scales, solute concentration at the drying interface tends to its equilibrium value, i.e. $\varphi_i \to \varphi^\star$, and the evaporation driving force asymptotically drops to zero. 
In this  regime, evaporation rates are small and the overall water transport is mainly dominated by diffusion within the liquid phase, as already known. 
In this paper we show that, for some complex fluids,  evaporation rates are humidity-insensitive in a wide range of humidity, unlike expected.

This result comes from two superimposed effects. First, the sharp decrease of the water chemical activity at high concentrations (as observed for instance for polymer solutions)  leads to small variations of $\varphi^\star$ along with $a_e$, and thus to small variations of the corresponding evaporation rate. 
This feature is specific to the case of molecular mixtures only, and we do not expect similar observations for colloidal dispersions for instance. Indeed, uni-directional drying of dispersions may also induce the formation of (porous) non-volatile aggregates at the drying interface, but they do not  
not in turn affect the chemical activity of the solvent, and hence evaporation rates, see e.g.~\cite{Lidon:14,Brown:02,Style:11}. 
In the case of molecular binary mixtures, the decrease of the mutual diffusion coefficient of the mixture at high concentration is key to explain why evaporation rates are remarkably independent of $a_e$ over a wide humidity range, see for instance Fig.~\ref{fig:CaseConstantDevap}. The basic mechanism is the following. The strong decrease of $D(\varphi)$ induces a steep concentration gradient at the drying interface. The concentration profile thus reaches values for which the mutual diffusion becomes again close to $D_0$, at positions $x$ close to the drying interface ($\varphi = \varphi_c$ at $x=x_c$ in our model with a piecewise constant $\hat{D}(\varphi)$). The remaining part of the profile, which contributes mainly to the value of the evaporation rate, is therefore shielded from the humidity variations which only play a role on this thin layer. This is the key interpretation provided by Roger {\it et al.} in
the context of water transport through self-assembled concentrated phases to explain why evaporation rates are  insensitive to $a_e$~\cite{Roger:16}.
Surprisingly, despite (i) an extensive survey of the abundant literature concerning notably the drying of polymeric films, and (ii) the simplicity of the above theoretical description, such a regime has never been discussed theoretically to the best of our knowledge.

Note that Roger {\it et al.} used the concept of {\it permeability} instead of mutual diffusion to interpret their result. Mutual diffusion actually results from an interplay between  a thermodynamic factor taking into account the variation
of the chemical activity along with the concentration, and a mobility factor related to the relative transport solutes/solvent~\cite{DeGrootMazur}. In the context of molecular mixtures, $D(\varphi)$ is often written as:
\begin{align} 
D(\varphi) = - M(\varphi) \frac{\partial \ln a(\varphi)}{\partial \varphi}\,,  \label{Dphi1}
\end{align}
where $M(\varphi)$ is a mobility factor, which takes  for instance the following form in the field of 
colloidal dispersions:
\begin{align} 
 M(\varphi) = - \varphi \frac{k}{\eta}\frac{k_B T}{v_m}\,,  \label{Dphi2}
\end{align}
where $k$ is the water permeability through the colloidal dispersion, $\eta$ the water viscosity, $T$ the absolute temperature, and $v_m$ the molecular volume of water, see e.g.~Ref.~\cite{Peppin:05}.
For self-assembled phases of surfactants, permeability (or equivalently the mobility factor) may decrease strongly within a phase owing to its specific texture~\cite{Spaar:01}. This may thus lead to the conclusion that the humidity-insensitive evaporation-driven water transport is related to self-assembly. 
 However, we showed in the present work that such a regime is also  expected for any binary mixture, independently of any phase transition,  when its mutual diffusion coefficient decreases at high concentration. 
Note finally that the decrease of $D(\varphi)$ may not be extremely strong, as a sharp decrease of the  water chemical activity also contributes significantly to this effect.

In the model investigated above, we have assumed local thermodynamic equilibrium conditions. Our model may thus 
apply to a wide range of experimental situations including for instance the drying of polymer solutions when no evaporation-induced glass transition occurs, or water transport through self-assembled phases of surfactants (without including the nucleation and growth steps of the latter). However our model,
and particularly the steady configuration shown in Fig.~\ref{fig:Sketch}(b), may also apply
to some extent to the description of one dimensional drying-induced water transport through 
cross-linked dilute hydrogels. One can indeed derive similar equations  providing  elastic contributions  to the water chemical activity and to the mutual diffusion coefficient are negligible,  disregarding non-Fickian mass transport, and taking into account properly the one dimensional swelling of the network, see for instance Ref.~\cite{Jeck:11}, and also Refs.~\cite{Bertrand2016,Okunozo:06,Okuzono:08} for some investigations of the general case. We hope in a near future to investigate such an issue in more details, and particularly the role of the humidity.
For non-negligible elastic contributions for instance, the phenomenology described above, particularly in Fig.~\ref{fig:Sketch}, may not apply, as elastic effects may induce the growth of 
a (permeable) gel  phase at a critical concentration~\cite{Okunozo:06,Okuzono:08}. We hope as 
a research perspective to evaluate the significance of these elastic effects with respect to
the mechanisms unveiled above.
 Such experimental cases may indeed be relevant for a wide range of experimental situations including for instance drying-induced water  transport through biomaterials or through  soft contact lenses~\cite{Fornasiero:08}.

\appendix 

\section{Unidirectional drying vs. thick film drying} \label{AppA}

\begin{figure}[ht]
\begin{center}
\includegraphics{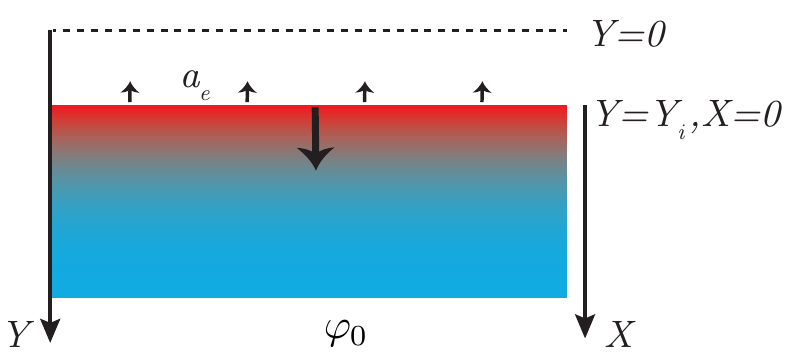}
\caption{Schematic view of the drying of a quiescent thick film.
\label{Thick}}
\end{center}
\end{figure}

We consider the 1D drying of a motionless thick
liquid film. The liquid gas interface
initially stands at height $Y=0$, then decreases
due to the volume loss induced by evaporation, see Fig.~\ref{Thick}.
 The solvent global mass balance reads:
\begin{eqnarray}
\frac{\text{d}Y_i}{\text{d}T} = J (a(\varphi_i) - a_e)\,, 
\end{eqnarray}
where $Y_i$ is the interface position, and $\varphi_i$ the solute concentration at the interface.
Within the volume, the solute volume fraction verifies:
\begin{eqnarray}
\partial_T \varphi = \partial_Y (D(\varphi) \partial_Y \varphi)~~\text{for}~~Y>Y_i\,.
\end{eqnarray}
We use the dimensionless units defined in Eq.~(\ref{eq:Varadim}) to get the following equations:
\begin{eqnarray}
&&\frac{\text{d}y_i}{\text{d}t} = a(\varphi_i) - a_e \,, \label{dyidt}\\
&&\partial_t \varphi = \partial_y (\hat{D}(\varphi) \partial_y \varphi)~~\text{for}~~y>y_i\,. \label{difffilm} 
\end{eqnarray}
Considering the motion of the liquid-gas interface at 
a velocity $\text{d}y_i/\text{d}t$, the non-volatility of the solute results in:
\begin{eqnarray}
\varphi_i\frac{\text{d}y_i}{\text{d}t} + (\hat{D}(\varphi) \partial_y \varphi)_{y=y_i} = 0\,.\label{nvfilm}
\end{eqnarray}
Assuming that the liquid film is thick enough to be considered as a semi-infinite medium yields  the second boundary condition: 
\begin{eqnarray}
\varphi(y\to \infty, t ) = \varphi_0.\label{bcfilm} 
\end{eqnarray}
The initial condition is:
\begin{eqnarray}
\varphi(y, t = 0) = \varphi_0~~\text{for}~~y>0.\label{initfilm}
\end{eqnarray}
One recovers strictly Eqs.~(\ref{eq:trans1}--\ref{eq:IC}) from the above model describing the drying of a quiescent thick film by using the change of variable:
\begin{eqnarray}
&&x = y(x,t)-y_i(t)\,, \label{eq:changevar}
\end{eqnarray}
and with the relation $v = \text{d}y_i/\text{d}t$.
These two problems are thus strictly equivalent.

\section{Numerical resolution of Eq.~(\ref{eq:steady1})} \label{AppRes}

Our aim is to compute the solution of Eq.~(\ref{eq:steady1}) for a fixed $\psi$ value given by Eq.~(\ref{eq:psi}), and with $v$ given by Eq.~(\ref{eq:trans2}). We proceeded as follows.
We used the solver ode113 (Matlab)
to compute the solutions of the first order ordinary differential equation Eq.~(\ref{eq:steady1})
for  a given humidity $a_e$, and for many different boundary conditions $\varphi(x=0)$  increasing in a logarithmic way from $\varphi_{j=1} = \varphi^\star - 0.5$ to $\varphi_{j=N} = \varphi^\star - \epsilon$ with $\epsilon = 10^{-5}$ and $N=100$ typically. 
To describe correctly the strong concentration gradient at the interface, 
each solution is calculated on a logarithmic $x$ scale.

For each solution with boundary condition $\varphi(x=0) = \varphi_{j}$, we then calculate
the total amount of solute $\psi_j$ using Eq.~(\ref{eq:psi}). The relation $\psi_j$ vs. $\varphi_{j}$ allows us to  estimate precisely using a linear interpolation, the boundary condition $\varphi(x=0)$ which leads to a given $\psi$ value, e.g. $\psi = 15$ for the cases shown in Fig.~\ref{fig:Steady}.

\section{Simple estimates of the onset of the falling rate period}\label{SoluceApp}
Our first aim is to derive an analytical approximation, at early time scales, of the solutions of Eqs.~(\ref{eq:trans1}-\ref{eq:trans2}) with boundary conditions given by Eqs.~(\ref{eq:BC1}-\ref{eq:BC2}) and initial condition Eq.~(\ref{eq:IC}). 

At early stages, concentration profiles follow $\varphi(x,t) \ll \varphi^\star$, $a(\varphi)\simeq 1$, $\hat{D}(\varphi) \simeq 1$,
and the evaporation rate given by Eq.~(\ref{eq:trans2}) is nearly constant, $v \simeq 1-a_e$. The 1D transport equation~Eq.~(\ref{eq:trans1}) can be thus written as
\begin{eqnarray}
&&\partial_t \varphi = - \partial_x j \label{eq:AppC1}\,,
\end{eqnarray}
with the non-dimensionalized solute flux $j = - (1-a_e)\varphi - \partial_x \varphi_x$.
Using Eq.~(\ref{eq:AppC1}), it is straightforward to demonstrate that the flux follows in turn
\begin{eqnarray}
&&\partial_t j = (1-a_e) \partial_x j +\partial_x^2 j  \label{eq:AppC1J}\,.
\end{eqnarray}
After a transient, the concentration process tends to an asymptotic regime for which the flux is stationary $\partial_t j = 0$, still assuming a dilute solution $\varphi(x,t) \ll \varphi^\star$. This steady flux can be 
estimated using Eq.~(\ref{eq:AppC1J}) and the boundary conditions Eqs.~(\ref{eq:BC1}-\ref{eq:BC2}) which  simply read $j(x=0) = 0$ and $j(x\gg 1) = - \varphi_0(1-a_e)$, leading finally to 
\begin{eqnarray}
j(x) =  - \varphi_0 (1-a_e) \left(1 - \exp\left[-(1-a_e)x\right]\right) \label{eq:AppC2}\,.
\end{eqnarray}

In this asymptotic regime, the rate of concentration given by Eq.~(\ref{eq:AppC1}) follows approximatively 
\begin{eqnarray}
\partial_t \varphi \simeq \varphi_0  (1-a_e)^2 \exp\left(-(1-a_e)x\right)\,,
\end{eqnarray}
and the concentration profiles are thus well-approximated by
\begin{eqnarray}
\varphi(x,t) \simeq \varphi_0  + t \, \varphi_0  (1-a_e)^2 \exp\left(-(1-a_e)x\right)\,,
\end{eqnarray}
using the initial condition~Eq.~(\ref{eq:IC}).
The onset of the falling rate period can be estimated by $\varphi(x=0,t_c) = \varphi^\star$ using the previous relation, leading to:
\begin{eqnarray}
t_c = \frac{\varphi^\star - \varphi_0}{(1-a_e)^2 \varphi_0}\,,
\end{eqnarray}
see the circles shown in Fig.~\ref{fig:CasePVA2}.

\section{Self-similar solutions} \label{AppB}
We use the model of evaporation from a thick film described in apprendix \ref{AppA} to find a self-similar solution in the long-time scale regime.
Eq.~(\ref{dyidt}) is replaced by the following Dirichlet boundary condition:
\begin{eqnarray}
\varphi(y_i,t) = \varphi^\star\,,\label{diricfilm}
\end{eqnarray}
where $\varphi^\star$ is a constant given by $a(\varphi^\star) = a_e$. This is fully justified as $\text{d} y_i / \text{d}t \to 0$ in the long time scale regime
(see Eq.~\ref{dyidt}).
Equations~(\ref{difffilm}--\ref{bcfilm}) and (\ref{diricfilm})
are rewritten assuming
that the solute volume fraction $\varphi(y,t)$ depends on a single variable $\tilde{u}(y,t)$. As done
classically for diffusion problems, we choose $\tilde{u}(y,t)=y/(2\sqrt{t})$~\cite{Crank}.
The partial derivatives equation~(\ref{difffilm}) turns to the ordinary differential equation: 
\begin{eqnarray}
\frac{\text{d}}{\text{d} \tilde{u}} \left[\hat{D}(\varphi) \frac{\text{d}\varphi}{\text{d} \tilde{u}} \right] + 2 \tilde{u} \frac{\text{d}\varphi}{\text{d} \tilde{u}} = 0~~\text{for}~~\tilde{u}>\tilde{u}_i \label{collapseSelf}
\end{eqnarray}
where $\tilde{u}_i = y_i//(2\sqrt{t})$.
Equations~(\ref{bcfilm}) and (\ref{initfilm}) collapse onto a single boundary condition:
\begin{eqnarray}
\varphi(\tilde{u}\to \infty) = \varphi_0. \label{collapseBC}
\end{eqnarray}
The Dirichlet condition~(\ref{diricfilm}) now reads:
\begin{eqnarray}
\varphi(\tilde{u}_i) = \varphi^\star\,.\label{diricfilm2}
\end{eqnarray}
As $\varphi^\star$ is a constant, a necessary condition for
the self-similar solution to exist is $\tilde{u}_i = \alpha$ where $\alpha$ is a constant to be determined.
Immediate consequences of this condition are:
\begin{eqnarray}
	y_i = 2\alpha \sqrt{t}~~\text{and}~~ \frac{\text{d}y_i}{\text{d} t} = \frac{\alpha}{\sqrt{t}}\,. \label{yalpha}
\end{eqnarray}
The boundary condition Eq.~(\ref{nvfilm}) along with Eq.~(\ref{yalpha}) yields:
\begin{eqnarray}
2 \alpha \varphi_i  + \hat{D}(\varphi_i) \frac{\text{d}\varphi}{\text{d} \tilde{u}}(\tilde{u}=\alpha) = 0\,, \label{self_bc}
\end{eqnarray}
whose unknown is $\alpha$.
The self-similar solution
exists if Eq.~(\ref{self_bc}) has a solution.

\subsection{Constant mutual diffusion coefficient $\hat{D}(\varphi) = 1$}
 	We solve Eqs.~(\ref{collapseSelf}--\ref{diricfilm2}, \ref{self_bc}) to find:
\begin{eqnarray}
\varphi(\tilde{u}) = \varphi_0 + (\varphi^\star-\varphi_0)\frac{\text{erfc}(\tilde{u})}{\text{erfc}(\alpha)}\,.\label{soluce}
\end{eqnarray}
Injecting Eq.~(\ref{soluce}) into (\ref{self_bc}) yields:
\begin{eqnarray}
\frac{\varphi^\star-\varphi_0}{\varphi^\star}=\sqrt{\pi} \; \alpha \; \text{erfc}(\alpha) \exp(\alpha^2)\,,\label{solucealpha}
\end{eqnarray}
hence leading to $\alpha$.

To recover the solution for the problem of unidirectional drying, we now use the change of variable~(\ref{eq:changevar})
and we define $u(x,t)=x/(2\sqrt{t})$. The self-similar solution is  now:
\begin{eqnarray}
\varphi(u) = \varphi_0 + (\varphi^\star-\varphi_0)\frac{\text{erfc}(u+\alpha)}{\text{erfc}(\alpha)}\,, \label{soluce2}
\end{eqnarray}
and Eq.~(\ref{solucealpha}) is unchanged.
With the assumption $\varphi_0 \ll \varphi^\star$, a simple approximate solution of Eq.~(\ref{solucealpha}) is obtained by a second order series expansion of erfc at $\alpha \to \infty$. We get:
\begin{eqnarray}
\alpha \simeq \sqrt{\frac{\varphi^\star}{2\varphi_0}}\,,
\end{eqnarray}
where the evaporation rate follows:
\begin{eqnarray} 
	v = \sqrt{\frac{\varphi^\star}{2\varphi_0 t}}\,.  \label{Vtrans}
\end{eqnarray}

\subsection{Piecewise constant mutual diffusion coefficient $\hat{D}(\varphi)$}
We consider now a piecewise constant diffusion coefficient, 
falling from 1 to a constant value $d$ at a solute volume fraction $\varphi_c$, see Eq.~(\ref{eq:piecewise}) and Fig.~\ref{fig:Da}(b). 
We define $y_c$ the abscissa such that $\varphi(y_c,t) = \varphi_c$.
With the change of variable $\tilde u(y,t)=y/(2 \sqrt{t})$, 
the solute volume fraction $\varphi(\tilde u)$ verifies Eq.~(\ref{collapseSelf}) with $\hat D(\varphi) = d$ for $\tilde u_i < \tilde u < \tilde u_c$,
and $\hat D(\varphi) = 1$ for $\tilde  u_c < \tilde  u$, where $\tilde  u_i = y_i/(2 \sqrt{t})$ and $\tilde  u_c = y_c/(2 \sqrt{t})$.
The existence of a self-similar solution requires $y_i = 2 \alpha \sqrt{t}$ and $y_c = 2 \beta \sqrt{t}$,
where $\alpha$ and $\beta$ are two constants to be determined.
Boundary conditions~(\ref{collapseBC}--\ref{diricfilm2}) are supplemented with flux conservation equations at $\tilde  u=\tilde  u_i=\alpha$ and $\tilde  u=\tilde  u_c=\beta$:
\begin{eqnarray} 
	&&2 \alpha \varphi^*  + d \frac{ \mathrm{d} \varphi}{\mathrm{d} \tilde  u}(\tilde  u=\alpha) = 0, \label{Self1} \\
	&&d \frac{\mathrm{d} \varphi}{\mathrm{d} \tilde  u}(\tilde  u=\beta^-) = \frac{\mathrm{d} \varphi}{\mathrm{d} \tilde  u}(\tilde  u=\beta^+). \label{Self2}
\end{eqnarray}
The solution is 
\begin{align} \label{SolSW1}
	\varphi(\tilde u) = \left ( \varphi^*-\varphi_c \right ) \frac {\erfc(\frac{\tilde  u}{\sqrt d}) - \erfc(\frac{\alpha}{\sqrt d})} {\erfc(\frac{\alpha}{\sqrt d})-\erfc(\frac{\beta}{\sqrt d})}  + \varphi^*\,, 
\end{align}
for $\tilde  u_i < \tilde  u < \tilde  u_c$, and
\begin{align} \label{SolSW2}
	\varphi(\tilde  u) = \left (\varphi_c-\varphi_0 \right ) \frac{\erfc(\tilde  u)} {\erfc(\beta)}  + \varphi_0\,,
\end{align}
for $\tilde  u_c < \tilde  u$. 
Injecting Eqs.~(\ref{SolSW1}--\ref{SolSW2}) in Eqs.~(\ref{Self1}--\ref{Self2}) provides two equations to be solved to get $\alpha$ and $\beta$:
\begin{align} 
	& \sqrt {\frac{\pi}{d}} \left [ \erfc(\frac{\alpha}{\sqrt d}) -  \erfc(\frac{\beta}{\sqrt d}) \right ] \exp (\frac{\alpha^2}{d})   \alpha   =	\frac{ \varphi^*-\varphi_c}{\varphi^*}\,, \label{AlphaEq} \\
	&  \frac {\left [ \erfc(\frac{\alpha}{\sqrt d}) -  \erfc(\frac{\beta}{\sqrt d}) \right ] \exp(\frac{\beta^2}{d})}  { \erfc (\beta) \sqrt {d} \exp{(\beta^2)}} =  \frac {\varphi^* - \varphi_c}{\varphi_c - \varphi_0}\,.  \label{BetaEq}
\end{align}

Eqs.~(\ref{AlphaEq}--\ref{BetaEq}) can be solved numerically.
However, we are particularly interested in the asymptotic case $d \to 0$, for which an analytical solution can be found.
With the approximation $\erfc(z \to \infty) \simeq  \exp(-z^2)/(\sqrt{\pi} z)$,  Eqs.~(\ref{AlphaEq}--\ref{BetaEq}) turn to:
\begin{eqnarray} 
&&	\frac{\alpha}{\beta} \exp \left ( \frac{\alpha^2 - \beta^2}{d} \right ) = \frac{\varphi_c}{\varphi^*}\,, \label{AlphaEqSer} \\
&&	\frac { \sqrt {\pi} \beta \erfc{(\beta)} \exp{(\beta^2)}} { \exp{  \left ( \frac{\beta^2 - \alpha^2}{d} \right ) \frac{\beta} {\alpha} -1}} = \frac{\varphi_c - \varphi_0}{\varphi^* - \varphi_c}\,.  \label{BetaEqSer}
\end{eqnarray}
Because Eqs.~(\ref{AlphaEqSer}--\ref{BetaEqSer}) require non zeros left hand sides to be satisfied, $d \to 0$ implies $\beta \to \alpha$. 
Using Eq.~(\ref{AlphaEqSer}) to eliminate $\alpha$ in Eq.~(\ref{BetaEqSer}) yields:
\begin{align} 
	\sqrt{\pi} \; \beta \erfc{(\beta)} \exp{(\beta^2)} = \frac {\varphi_c - \varphi_0} {\varphi_c}  \label{BetaEqSer2}.
\end{align}
By analogy with Eq.~(\ref{solucealpha}), and considering the assumption $\varphi_0 \ll  \varphi_c$, we get finally:
\begin{align} 
	\alpha \simeq \beta \simeq \sqrt{ \frac{\varphi_c}{2 \varphi_0} } .  \label{SolD2}
\end{align}

\begin{acknowledgments}
The authors thank ANR EVAPEC (13-BS09-0010-01)
for funding.
\end{acknowledgments}


\end{document}